\documentclass[twocolumn]{iopart}
\usepackage{iopams}
\usepackage{graphicx,epsf}
\usepackage{color}
\usepackage{mathrsfs}
\usepackage{dcolumn}
\usepackage{cite}

\begin{document}

\title[]{Gyrokinetic theory of the nonlinear saturation of toroidal Alfv\'en eigenmode}

\author{Zhiyong Qiu$^{1}$, Liu Chen$^{1, 2}$ and Fulvio Zonca$^{3, 1}$}


\address{$^1$Institute for    Fusion Theory and Simulation and Department of Physics, Zhejiang University, Hangzhou, P.R.C}
\address{$^2$Department of   Physics and Astronomy,  University of California, Irvine CA 92697-4575, U.S.A.}
\address{$^3$ ENEA, Fusion and Nuclear Safety Department, C. R. Frascati, Via E. Fermi 45, 00044 Frascati (Roma), Italy}

\begin{abstract}

{
Nonlinear saturation of toroidal Alfv\'en eigenmode (TAE) via ion induced scatterings is investigated in the short-wavelength gyrokinetic regime. It is found that the nonlinear evolution depends on the thermal ion $\beta$ value. Here, $\beta$ is the plasma thermal to magnetic pressure ratio. Both the saturation levels and associated energetic-particle transport coefficients are derived and estimated correspondingly.
}
\end{abstract}

\maketitle

\section{Introduction}\label{sec:intro}

In    burning plasmas  of  next generation devices such as ITER \cite{KTomabechiNF1991},    energetic particles (EP), e.g. fusion-alpha particles,  contribute  significantly to the total power density   and, consequently, could drive shear Alfv\'en wave (SAW) instabilities  \cite{YKolesnichenkoVAE1967,AMikhailovskiiSPJ1975,MRosenbluthPRL1975,LChenPoP1994,LChenRMP2016}.   SAW instabilities, in turn, can lead to enhanced EP  transport, degradation of plasma performance and, possibly, damaging of plasma facing components  \cite{IPBNF1999,AFasoliNF2007,RDingNF2015}.
Due to equilibrium magnetic field geometries and plasma nonuniformities,  SAW instabilities manifest themselves as EP continuum modes (EPM)  \cite{LChenPoP1994}  and/or various discretized  Alfv\'en eigenmodes (AE); e.g., the well-known toroidal Alfv\'en eigenmode (TAE) \cite{CZChengAP1985}.  The EP   anomalous transport rate is related to TAE amplitude and spectrum \cite{LChenJGR1999}, and thus, in-depth understanding of the nonlinear dynamics of TAE  is crucial for assessing the performance of  future burning plasmas   \cite{LChenRMP2016,IPBNF1999,AFasoliNF2007}.

Most  numerical investigations on TAE nonlinear dynamics focused on EP phase space dynamics induced by a single-toroidal-mode-number TAE  \cite{YTodoPoP1995,JLangPoP2010,JZhuPoP2013,SBriguglioPoP2014,JZhuNF2014}. There are some literatures  on  the effects of   mode couplings on the   TAE nonlinear dynamics   \cite{DSpongPoP1994,TSHahmPRL1995,FZoncaPRL1995,LChenPPCF1998,YTodoNF2010,LChenPRL2012,ZQiuEPL2013,ZQiuPoP2016,ZQiuNF2017}.   Hahm et al. \cite{TSHahmPRL1995} studied the TAE downward spectral cascading and eventually saturation induced by nonlinear ion Compton scattering. The nonlinearly saturated spectrum and overall electromagnetic perturbation amplitude are derived, and the resulting bulk ion heating rate was also obtained  in a later publication \cite{TSHahmPST2015}.
References   \citenum{FZoncaPRL1995} and \citenum{LChenPPCF1998}, meanwhile, demonstrated and analyzed TAE saturation via enhanced continuum damping due to nonlinearly narrowed SAW continuum gap.
The spontaneous excitation of axisymmetric zero frequency zonal structures (ZFZS) via modulational instability  is investigated in Ref. \citenum{LChenPRL2012};  and further extended to include the important effects of resonant EPs \cite{ZQiuPoP2016} and the   fine-scale  radial structures \cite{ZQiuNF2017}.  Recently, a new decay channel of TAE into a geodesic acoustic mode (GAM) and a kinetic TAE (KTAE) is proposed and analyzed \cite{ZQiuPRL2018}. It is shown that  this nonlinear decay process can lead to effective TAE saturation  and thermal ion heating via GAM Landau damping; i.e., an effective $\alpha$-channeling process  \cite{NFischPRL1992,NFischNF1994}. All the various nonlinear processes described so far may play similar important roles in situations of practical interest, depending on the plasma parameters. This makes the analysis complicated, since the various processes must be accounted for on the same footing. Furthermore, plasma conditions in present day machines and next generation devices are different, and correspond to different dominant nonlinear processes that must be considered in practical applications. Clarifying these issues is one aim of the present work and will be addressed in the following.

The theory presented in Ref. \citenum{TSHahmPRL1995}  considered that there exists many TAEs ($O(n^2q)$, with  $n$ being the characteristic toroidal mode number of most unstable TAE and $q$ the safety factor),  located at different radial positions with    slightly shifted frequencies due to local equilibrium parameters.  Furthermore, a low-$\beta$ regime was assumed, i.e., $\beta\ll \epsilon^2$, such that, in each triad interaction, a pump TAE decay into another TAE within the toroidicity induced SAW continuum gap and an electrostatic fluctuation near the  ion sound wave  (ISW) frequency range. Here, $\beta$ is the plasma thermal to magnetic pressure ratio,  and $\epsilon\equiv r/R_0$ is the inverse aspect ratio, with $r$ and $R_0$ being the minor and major radii of the torus.  More specifically, we note that TAEs are characterized by parallel wavenumber $|k_{\parallel}|\simeq 1/(2qR_0)$, and, thus, two counter-propagating TAEs  with radially overlapped mode structures  can couple  and generate an ISW fluctuation  with a much lower frequency and  $|k_{\parallel}|\simeq 1/(qR_0)$.  As the TAEs cascades toward lower frequencies, the wave energy is, eventually, absorbed via enhanced continuum damping near the lower SAW accumulation point.  The original  theory \cite{TSHahmPRL1995} adopted the nonlinear drift kinetic approach and considered the long wavelength MHD limit with $\omega/\Omega_{ci}\gg k^2_{\perp}\rho^2_i$ \footnote{To be more precise, $\omega/\Omega_{ci}\gg k_rk_{\theta}\rho^2_i$, as we shown in Sec. \ref{sec:parametric_instability}.}, which corresponds to $(T_i/T_E)/(q^2\epsilon )\ll \omega/\Omega_{ci}$ for TAEs excited by well-circulating EPs. The corresponding nonlinear couplings are due to the parallel  ponderomotive force from the  $\mathbf{\hat{b}}\cdot\delta \mathbf{J}\times \delta\mathbf{B}$ nonlinearity. Here, $\Omega_{ci}$ is the ion gyro-frequency, $k_{\perp}$ is the perpendicular wavenumber, $\rho_i$ is the ion Larmor radius,  $T_E$ and $T_i$ are the EP and bulk ion temperature, $\mathbf{\hat{b}}$ is the unit  vector along the equilibrium magnetic field, and $\delta \mathbf{J}$ and $\delta \mathbf{B}$ are the perturbed TAE current and magnetic field, respectively.  In next generation devices and plasmas of fusion interest, however, plasma parameters are such that the short wavelength   $k^2_{\perp}\rho^2_{i}>\omega/\Omega_{ci}$ regime applies \cite{LChenEPL2011,LChenRMP2016}, and,   one needs, instead, to adopt the nonlinear gyrokinetic approach \cite{LChenEPL2011}. This consideration is the primary motivation for the present analysis.

In the present work, we generalize the drift-kinetic theory of TAE saturation via ion induced scattering \cite{TSHahmPRL1995} to the fusion plasma  relevant short wavelength regime using nonlinear gyrokinetic theory \cite{EFriemanPoF1982}.  Both low- and high- $\beta$ regimes are considered.  In the lower $\beta$ limit,  our  analysis, following closely that of Ref. \citenum{TSHahmPRL1995}, shows that  in the gyrokientic regime,    the   nonlinear coupling coefficients are much bigger than those predicted in Ref. \citenum{TSHahmPRL1995}. As a consequence, our theory predicts lower levels of     TAE saturation   and EP transport.  In the higher-$\beta$ limit, since the   nonlinear coupling is maximized for ISW frequency larger than the TAE frequency mismatch with the SAW continuum,  the physics picture becomes different. That is, the TAE decays directly into an ISW fluctuation and a propagating lower kinetic TAE (LKTAE) \cite{FZoncaPoP1996}. Expressions for the nonlinear saturation levels in both regimes are derived, and then applied to estimate the corresponding EP transport rates for future burning plasmas.

The rest of the paper is organized as follows. In Sec. \ref{sec:model}, the theoretical model is given, which is then used to derive the nonlinear parametric dispersion relation in Sec. \ref{sec:nl_dr}.  The nonlinear TAE spectrum evolution  and the  saturation level in the lower-$\beta$ limit are derived in Sec. \ref{sec:spectrum_evolution}.  The TAE saturation level in the higher-$\beta$ limit, meanwhile, is derived in  Sec. \ref{sec:high_beta_saturation}.  Corresponding EP transport rates in both   regimes  are evaluated in Sec. \ref{sec:transport} based on the quasilinear approach.  Finally, a conclusion is given in Sec. \ref{sec:summary}.

\section{Theoretical model}\label{sec:model}

To investigate the nonlinear TAE spectrum evolution, we adopt the standard nonlinear perturbation theory, and consider a
 toroidal Alfv\'en mode (TAM)\cite{FZoncaPoP1996}  $\mathbf{\Omega}_0=(\omega_0$, $\mathbf{k}_0)$   interacting with    another  TAM     $\mathbf{\Omega}_1=(\omega_1$, $\mathbf{k}_1)$ and generating   an   ISW fluctuation  $\mathbf{\Omega}_S=(\omega_S$, $\mathbf{k}_S)$. Here, TAMs represent  SAW instabilities in the TAE frequency range, strongly affected by toroidal effects \cite{LChenPS1995,FZoncaPoP1996}, including TAE, KTAE, as well as EPM.   Thus, the nonlinear equations derived in Sec. \ref{sec:nl_dr} can be applied to both TAE spectral energy transfer in the lower-$\beta$ limit, where $\mathbf{\Omega}_0$ and $\mathbf{\Omega}_1$ correspond, respectively, to test and background TAEs;  and TAE decaying into LKTAE in the higher-$\beta$ limit, where $\mathbf{\Omega}_1$ and $\mathbf{\Omega}_0$ correspond, respectively, to the pump TAE and LKTAE sideband\footnote{Note that, in this work, $\mathbf{\Omega}_1$ and $\mathbf{\Omega}_0$ correspond to the pump and sideband waves, contrary to usual notations.}.    The scalar potential $\delta \phi$ and parallel vector potential $\delta A_{\parallel}$ are used as the field variables, and one   has, $\delta\phi=\delta\phi_0+\delta\phi_1+\delta\phi_S$, with the subscripts $0$, $1$ and $S$ denoting  $\mathbf{\Omega}_0$,  $\mathbf{\Omega}_1$ and $\mathbf{\Omega}_S$, respectively.  Furthermore, $\delta\psi\equiv \omega\delta A_{\parallel}/(ck_{\parallel})$ is introduced as an alternative field variable, and one has $\delta\psi=\delta\phi$ in the ideal MHD limit.  Without loss of generality,  $\mathbf{\Omega}_0=\mathbf{\Omega}_1+\mathbf{\Omega}_S$ is adopted as the frequency/wavenumber matching  condition. For effective spectral transfer by ion induced scattering,  we have $|\omega_S|\sim O(v_{it}/qR_0)$; i.e., the ISW fluctuation frequency is comparable to thermal ion transit frequency.
Therefore,    $\mathbf{\Omega}_0$ and $\mathbf{\Omega}_1$ are counter-propagating  TAMs,  with $\omega_0\simeq\omega_1$ and $k_{\parallel,0}\simeq -k_{\parallel,1}$. Here,  $k_{\parallel}\equiv (nq-m)/(qR_0)$ is the   wavenumber parallel to equilibrium magnetic field.

For  high-$n$  TAMs, we adopt the following ballooning mode representation in the   $(r,\theta,\phi)$ field-aligned flux coordinates \cite{JConnorPRL1978}
\begin{eqnarray}
\delta\phi_0&=&A_0e^{i(n_0\phi-\hat{m}_0\theta-\omega_0t)}\sum_j e^{-ij\theta}\Phi_0(x-j)+c.c.,\nonumber\\
\delta\phi_1&=&A_1e^{i(n_1\phi-\hat{m}_1\theta-\omega_1t)}\sum_j e^{-ij\theta}\Phi_1(x-j+\delta_1)+c.c..\nonumber
\end{eqnarray}
Here, $m=\hat{m}+j$ with  $\hat{m}$ being the reference poloidal mode number, $x=n_0q-\hat{m}_0\simeq n_0q'(r_0)(r-r_0)$,
$r_0$ is the TAE localization position with $|n_0q(r_0) -\hat{m}_0|\simeq1/2$,   $\Phi$ is the fine radial structure associated with $k_{\parallel}$ and magnetic shear, $\delta_1\equiv (n_1-n_0)q+\hat{m}_0-\hat{m}_1\mp1$ is a small normalized radial shift accounting for possible misalignment of TAM radial mode structure \footnote{The $\mp$ sign is to be chosen according to the leading order value of $(n_1-n_0)q+\hat{m}_0-\hat{m}_1$ being $\pm1$ according to the parallel wave number matching condition.}, and $A$ is the mode amplitude. The ISW $\mathbf{\Omega}_S$,   on the other hand, can be written as
\begin{eqnarray}
\delta\phi_S=A_Se^{i(n_S\phi-m_S\theta-\omega_St)}\Phi_S.\nonumber
\end{eqnarray}
 $\Phi_S$ is determined  by $\Phi_0$ and $\Phi_1$ \cite{ZQiuNF2016}. Noting that,  for   ISM, the corresponding typical distance between mode rational surfaces is much wider than that of TAEs, i.e., $1/|n_Sq'(r_S)|\gg 1/|n_0q'(r_0)|, 1/|n_1q'(r_1)|$ as noted earlier $n_S\ll n_0, n_1$, we typically have $|n_Sq(r_S)-m_S|\simeq 1$ and $|\delta_1|\ll 1$.

The governing equations describing the nonlinear interactions among $\mathbf{\Omega}_0$, $\mathbf{\Omega}_1$ and  $\mathbf{\Omega}_S$, can then be derived from quasi-neutrality condition
\begin{eqnarray}
\frac{n_0e^2}{T_i}\left(1+\frac{T_i}{T_e}\right)\delta\phi_k=\sum_s\left\langle q J_k\delta H_k \right\rangle_s,\label{eq:QN}
\end{eqnarray}
and nonlinear gyrokinetic vorticity equation
\begin{eqnarray}
&&\frac{c^2}{4\pi \omega^2_k}B\frac{\partial}{\partial l}\frac{k^2_{\perp}}{B}\frac{\partial}{\partial l}\delta \psi_k +\frac{e^2}{T_i}\left\langle (1-J^2_ k)F_0\right\rangle\delta\phi_k\nonumber\\
&&-\sum_s\left\langle\frac{q}{\omega_k}J_k\omega_d\delta H_k \right\rangle_s\nonumber\\
&=&-i\frac{c}{B_0\omega_k}\sum_{\mathbf{k}=\mathbf{k}'+\mathbf{k}''} \mathbf{\hat{b}}\cdot\mathbf{k}''\times\mathbf{k}'\left [ \frac{c^2}{4\pi}k''^2_{\perp} \frac{\partial_l\delta\psi_{k'}\partial_l\delta\psi_{k''}}{\omega_{k'}\omega_{k''}} \right.\nonumber\\
&&\left.+ \left\langle e(J_kJ_{k'}-J_{k''})\delta L_{k'}\delta H_{k''}\right\rangle \right].
\label{eq:vorticityequation}
\end{eqnarray}
Here, $J_k\equiv J_0(k_{\perp}\rho)$ with $J_0$ being the Bessel function of zero index, $\rho=v_{\perp}/\Omega_c$, $\Omega_c$ is the cyclotron frequency, $F_0$ is the equilibrium particle distribution function,  $\sum_s$ is the summation on different particle species,    $\omega_d=(v^2_{\perp}+2
v^2_{\parallel})/(2 \Omega_c R_0)\left(k_r\sin\theta+k_{\theta}\cos\theta\right)$ is the magnetic drift frequency,  $l$ is the length along the equilibrium magnetic field line, $\delta L_k\equiv\delta\phi_k-k_{\parallel} v_{\parallel}\delta\psi_k/\omega_k$; and other notations are standard.  The dominant  nonlinear terms in the vorticity equation are        Maxwell and Reynolds stresses; i.e., the first and second terms on the right hand side of equation (\ref{eq:vorticityequation}), respectively\footnote{More precisely, the Reynolds stress is recovered from the long wavelength limit of the second term on the right hand side of equation (\ref{eq:vorticityequation})\cite{LChenRMP2016}.}.  Furthermore, $\langle\cdots\rangle$ indicates velocity space integration and $\delta H$ is the nonadiabatic particle response, which can be derived from nonlinear gyrokinetic equation \cite{EFriemanPoF1982}:
\begin{eqnarray}
&&\left(-i\omega+v_{\parallel}\partial_l+i\omega_d\right)\delta H_k=-i\omega_k\frac{q}{m}QF_0J_k\delta L_k \nonumber\\
&&\hspace*{4em}-\frac{c}{B_0}\sum_{\mathbf{k}=\mathbf{k}'+\mathbf{k}''}\mathbf{\hat{b}}\cdot\mathbf{k''}\times\mathbf{k'}J_{k'}\delta L_{k'}\delta H_{k''}\label{eq:NLGKE}.
\end{eqnarray}
Here, $QF_0=(\omega\partial_E-\omega_*)F_0$ with $E=v^2/2$, $\omega_*F_0=\mathbf{k}\cdot\mathbf{\hat{b}}\times\nabla F_0/\Omega_c$ is related to the expansion free energy. In this work, we assume that the TAE drive is from EPs while neglect the pressure gradient of bulk plasmas. On the other hand,  for EPs we assume    $QF_{0,E}\simeq  -\omega_{*,E} F_{0,E}$,  for EP drive strong enough to drive TAE unstable \cite{GFuPoFB1989}.

\section{Parametric decay instability}\label{sec:nl_dr}

In this section,  the   equations  describing the nonlinear evolution of  $\mathbf{\Omega}_0$  due to interactions with  $\mathbf{\Omega}_1$   are derived.  They are very closely related to those of Ref. \citenum{LChenEPL2011} for parametric decay of KAWs in uniform plasmas, with differences due to  the peculiar features associated with the toroidal geometry. The derivation  follows the standard procedure of a nonlinear perturbation theory. At the leading order, linear particle responses to  TAMs  and ISW are derived, which are then used at the next order in the small amplitude expansion to derive the  nonlinear equations describing ion sound mode generation by beating of  $\mathbf{\Omega}_0$ and $\mathbf{\Omega}_1$. Finally, the  equations describing  nonlinear evolution of  $\mathbf{\Omega}_0$  due to   the $\mathbf{\Omega}_S$ and  $\mathbf{\Omega}_1$  coupling  is derived.

Separating   $\delta H_{k,s}=\delta H^L_{k,s}+\delta H^{NL}_{k,s}$,  the linear particle responses   to the   electrostatic ISW  can be derived noting the $\omega_S\sim  O(v_{i}/(qR_0))$ ordering for effective ion induced scattering. One obtains
\begin{eqnarray}
\delta H^L_{S,e}&=&0,\label{eq:linear_electron_ISM} \\
\delta H^L_{S,i}&=&\frac{e}{T_i}F_0\frac{\omega_S}{\omega_S-k_{\parallel,S}v_{\parallel}}J_S\delta\phi_S. \label{eq:linear_ion_ISM}
\end{eqnarray}
Small magnetic drift orbit width ordering  ($|\omega_{d,S}|\ll |v_{it}/(qR_0)|$) and $|\omega_{*,S}|\ll|\omega_S|$ are also used here.  Linear particles responses to the high-$n$  TAMs  can also be derived noting the $k_{\parallel,T}v_{te}\gg\omega_T\gg k_{\parallel,T}v_{i}\gg\omega_{d,i},\omega_{d,e}$ ordering. At the leading order one obtains
\begin{eqnarray}
\delta H^L_{T,e}&=&-\frac{e}{T_e}F_0\delta\psi_T,\label{eq:linear_electron_TAM} \\
\delta H^L_{T,i}&=&\frac{e}{T_i}F_0J_T\delta\phi_T. \label{eq:linear_ion_TAM}
\end{eqnarray}
Equations (\ref{eq:linear_electron_ISM}) to (\ref{eq:linear_ion_TAM}) are used below in the nonlinear analysis at the next order in the small amplitude expansion.

\subsection{Nonlinear ion sound wave fluctuation generation}

The predominantly electrostatic ISW fluctuation generation can be derived from quasi-neutrality condition, with the  nonlinear particle responses  derived from nonlinear gyrokinetic equation. For electrons with  $k_{\parallel,S}v_{e}\gg\omega_S,\omega_{d,S}$,   the nonlinear gyrokinetic equation becomes
\begin{eqnarray}
v_{\parallel}\partial_l\delta H^{NL}_{S,e}&=&-\frac{c}{B_0}\sum\hat{\mathbf{b}}\cdot\mathbf{k''}\times\mathbf{k'}\delta L_{k'}\delta H_{k'',e}\nonumber\\
&\simeq&-\hat{\Lambda}\frac{e}{T_e}F_0v_{\parallel}\left(\frac{k_{\parallel,1^*}}{\omega_{1^*}}-\frac{k_{\parallel,0}}{\omega_0}\right)\delta\phi_0\delta\psi_{1^*},\nonumber
\end{eqnarray}
with $\hat{\Lambda}\equiv (c/B_0)\hat{\mathbf{b}}\cdot\mathbf{k_0}\times\mathbf{k_{1^*}}$, and the superscript ``$*$" in the subscripts denoting the corresponding quantity of the complex conjugate component. Noting that $\omega_{1^*}\simeq-\omega_0$, $k_{\parallel,1^*}\simeq k_{\parallel,0}$, and that $k_{\parallel,S}\simeq 2k_{\parallel,0}$, one then has
\begin{eqnarray}
\delta H^{NL}_{S,e}=-i\frac{\hat{\Lambda}}{\omega_0}\frac{e}{T_e}F_0\delta\psi_0\delta\psi_{1^*}.
\end{eqnarray}

Nonlinear ion response to $\mathbf{\Omega}_S$, on the other hand, can be derived,  noting the $\omega_S\sim k_{\parallel,S}v_{it}\gg\omega_{d,S}$ ordering
\begin{eqnarray}
\delta H^{NL}_{S,i}=-i\frac{\hat{\Lambda}}{\omega_0}\frac{e}{T_i}F_0\frac{k_{\parallel,S}v_{\parallel}}{\omega_S-k_{\parallel,S}v_{\parallel}}J_0J_1\delta\phi_0\delta\phi_{1^*}.
\end{eqnarray}

Substituting $\delta H_{S,i}$ and $\delta H_{S,e}$ into quasi-neutrality condition, one then obtains the nonlinear $\mathbf{\Omega}_S$ equation
\begin{eqnarray}
\mathscr{E}_S\delta\phi_S=i\frac{\hat{\Lambda}}{\omega_0}\beta_1\delta\phi_0\delta\phi_{1^*}.\label{eq:NL_quasimode}
\end{eqnarray}
Here,
$\mathscr{E}_S\equiv 1+\tau+\tau\Gamma_S\xi_SZ(\xi_S)$ is the linear dispersion function of $\mathbf{\Omega}_S$, with $\tau\equiv T_e/T_i$, $\Gamma_S\equiv\langle J^2_SF_0/n_0\rangle$, $\xi_S\equiv\omega_S/(k_{\parallel,S}v_{it})$ and $Z(\xi_S)$ being the well known plasma dispersion function, defined as
\begin{eqnarray}
Z(\xi_S)\equiv\frac{1}{\sqrt{\pi}}\int^{\infty}_{-\infty}\frac{e^{-y^2}}{y-\xi_S}dy.\nonumber
\end{eqnarray}
Furthermore,  $\beta_1\equiv \sigma_0\sigma_{1}+\tau\hat{F}_1\left(1+\xi_SZ(\xi_S)\right)$,  with $\hat{F_1}\equiv\langle J_0J_1J_SF_0/n_0\rangle$, $\sigma_k\equiv 1+\tau-\tau\Gamma_k$,  and $\sigma_k\neq1$ corresponding  to breaking of ideal MHD constraint and generation of  finite parallel electric field by kinetic effects, which is typically  not important for TAEs  in the SAW continuum gap, while, on the other hand, is crucial for LKTAEs. The $\sigma_T$'s, with the subscript ``T" denoting TAMs, are systematically kept in this paper to be consistent with the notations of Ref. \citenum{LChenEPL2011}, where the equations presented in Sec. \ref{sec:nl_dr} are originally derived for parametric decay of KAWs with arbitrary perpendicular wavenumber in uniform plasmas.

\subsection{Nonlinear      TAM  equations}

Nonlinear particle responses to    $\mathbf{\Omega}_0$, can be derived similarly. Noting  that $\mathbf{\Omega}_S$ could be heavily  ion Landau damped, the linear behavior $\sim\delta\phi_S$ can be of the same order of the formally nonlinear response $\sim\delta\phi_0\delta\phi_{1^*}$. Thus, one needs to include both  linear and nonlinear responses while deriving the nonlinear particle responses to $\mathbf{\Omega}_0$, which can be readily  derived as \cite{LChenEPL2011}\footnote{Please, note the slightly different normalization used here and in Ref. \citenum{LChenEPL2011}, mostly connected with the definition of $\hat{\Lambda}$.}
\begin{eqnarray}
\delta H^{NL}_{0,e}&=&-\frac{\hat{\Lambda}^2}{\omega^2_0}\frac{e}{T_e}F_0\sigma^2_1\sigma_0|\delta\phi_1|^2\delta\phi_0,\label{eq:NL_electron_test}\\
\delta H^{NL}_{0,i}&=&i\frac{\hat{\Lambda}}{\omega_0}\frac{e}{T_i}F_0\frac{k_{\parallel,S}v_{\parallel}}{\omega_S-k_{\parallel,S}v_{\parallel}}\left[J_1J_S\delta\phi_{S}\delta\phi_1\right.\nonumber\\
&&\left.-i(\hat{\Lambda}/\omega_0)J^2_1J_0|\delta\phi_1|^2\delta\phi_0\right].\label{eq:NL_ion_test}
\end{eqnarray}

Substituting equations (\ref{eq:NL_electron_test}) and (\ref{eq:NL_ion_test}) into the quasi-neutrality condition, equation (\ref{eq:QN}), one has
\begin{eqnarray}
\delta\psi_0=\left(\sigma_0+\sigma^{(2)}_0\right)\delta\phi_0+D_0\delta\phi_1\delta\phi_{S},\label{eq:NL_QN_test}
\end{eqnarray}
in which,
\begin{eqnarray}
\sigma^{(2)}_0&\equiv&\hat{\Lambda}^2\left[-\sigma^2_1\sigma_0+\tau\hat{F}_2\left(1+\xi_SZ(\xi_S)\right)\right] |\delta\phi_1|^2/\omega^2_0,\nonumber\\
D_0&\equiv& i\hat{\Lambda}\tau\hat{F}_1\left[1+\xi_S Z(\xi_S)\right]/\omega_0,\nonumber\\
\hat{F}_2&\equiv& \left\langle J^2_0J^2_1(F_0/n_0)\right\rangle.\nonumber
\end{eqnarray}

The nonlinear  vorticity equation of $\mathbf{\Omega}_0$, is
\begin{eqnarray}
\left[\frac{1-\Gamma_0+\alpha^{(2)}_0/\omega^2_0}{b_0}\delta\phi_0-\frac{k^2_{\parallel,0}V^2_A}{\omega^2_0}\delta\psi_0\right]=\frac{D_2}{b_0}\delta\phi_1\delta\phi_{S},\label{eq:NL_vorticity_test}
\end{eqnarray}
with
\begin{eqnarray}
\alpha^{(2)}_0&=& \hat{\Lambda}^2 \left(\hat{F}_2-\hat{F}_1\right)\left(1+\xi_SZ(\xi_S)\right) |\delta\phi_1|^2,\nonumber\\
D_2&=&-i \hat{\Lambda}\left[\hat{F}_1(1+\xi_SZ(\xi_S))-\Gamma_S\xi_SZ(\xi_S)-\Gamma_1\right]/{\omega_0}.\nonumber
\end{eqnarray}

Substituting equation (\ref{eq:NL_QN_test}) into equation (\ref{eq:NL_vorticity_test}), one then obtains the nonlinear eigenmode equation of $\mathbf{\Omega}_0$:
\begin{eqnarray}
&&\left(\mathscr{E}_0+ \mathscr{E}^{NL}_0\right)\delta\phi_0=- \left(D_2 \omega^2_0/b_0 +   k^2_{\parallel,0}V^2_A D_0\right)\delta\phi_1\delta\phi_{S}. \nonumber\\
&& \label{eq:NL_test_DR_WKB}
\end{eqnarray}
Here, $\mathscr{E}_0\equiv\mathscr{E}_T(\mathbf{\Omega}_0)$ is the linearized wave operator  of   $\mathbf{\Omega}_0$ \cite{CZChengAP1985,FZoncaPoP1996,FZoncaPoFB1993,FZoncaPoP2014b}, with $\mathscr{E}_T$ defined as
$\mathscr{E}_T\equiv k^2_{\parallel,T}V^2_A \sigma_T-(1-\Gamma_T)\omega^2_T/b_T$, and $\mathscr{E}^{NL}_0\equiv - \alpha^{(2)}_0/b_0+k^2_{\parallel,0}V^2_A\sigma^{(2)}_0$.
The TAM eigenmode  dispersion relation can then be derived noting the $V^2_A\propto 1-2\epsilon_0\cos\theta$ dependence on poloidal angle $\theta$ \cite{FZoncaPoP1996,FZoncaPoFB1993,CZChengAP1985,FZoncaPoP2014b} with $\epsilon_0=2(r/R_0+\Delta')$ and $\Delta'$ being Shafranov shift. Meanwhile,  $\sigma^{(2)}$ and $\alpha^{(2)}$ correspond, respectively, to the contribution of  nonlinear particle response to ISW on  ideal MHD constraint breaking  and Reynolds stress.

The right hand side of equation (\ref{eq:NL_test_DR_WKB}) can be simplified using the expressions of $\mathscr{E}_k$ and $\sigma_k$, and one has
\begin{eqnarray}
\left( \mathscr{E}_0+ \mathscr{E}^{NL}_0\right)\delta\phi_0= i\frac{\omega_0\hat{\Lambda}\beta_2}{b_0\tau} \delta\phi_1\delta\phi_{S}, \label{eq:NL_test_DR_WKB_2}
\end{eqnarray}
with $\beta_2\equiv\beta_1/\sigma_0-\mathscr{E}_S$.
Substituting equation (\ref{eq:NL_quasimode}) into (\ref{eq:NL_test_DR_WKB_2}), we obtain
\begin{eqnarray}
\left( \mathscr{E}_0+ \mathscr{E}^{NL}_0\right)\delta\phi_0=  -\frac{\hat{\Lambda}^2\beta_1\beta_2}{b_0\tau\mathscr{E}_S}|\delta\phi_1|^2\delta\phi_0.\label{eq:NL_test_DR_3}
\end{eqnarray}
Equation (\ref{eq:NL_test_DR_3}) describes the nonlinear evolution of   $\mathbf{\Omega}_0$ due to the nonlinear interactions with $\mathbf{\Omega}_1$. Ion Compton scattering related to  ion Landau damping of the ISW fluctuation may play an important role  for TAE saturation, and consistently,  we re-write the coefficients explicitly as  functions of  $\mathscr{E}_S$:
\begin{eqnarray}
\mathscr{E}^{NL}_0=- \frac{\hat{\Lambda}^2}{b_0}|\delta\phi_1|^2\left(\hat{G}_1+\hat{G}_2\mathscr{E}_S\right),\nonumber
\end{eqnarray}
with
\begin{eqnarray}
\hat{G}_1&=&(1-\Gamma_0)\sigma^2_1- \sigma_S\hat{G}_2, \nonumber\\
\hat{G}_2&=&\left(\hat{F}_2-\hat{F}_1-(1-\Gamma_0)\tau\hat{F}_2/\sigma_0\right)/(\tau\Gamma_S).\nonumber
\end{eqnarray}
 On the other hand,
\begin{eqnarray}
\frac{\beta_1\beta_2}{\tau\mathscr{E}_S}=\hat{H}_1+\hat{H}_2\mathscr{E}_S+\frac{\hat{H}_3}{\mathscr{E}_S},\nonumber
\end{eqnarray}
with
\begin{eqnarray}
\hat{H}_1&=& \left(\sigma_0\sigma_1-\hat{F}_1\sigma_S/\Gamma_S\right)\left(2\hat{F}_1/\Gamma_S-\sigma_0\right)/(\tau\sigma_0),\nonumber\\
\hat{H}_2&=& \hat{F}_1 \left(\hat{F}_1/\Gamma_S-\sigma_0\right)/(\tau\sigma_0\Gamma_S),\nonumber\\
\hat{H}_3&=&  \left(\sigma_0\sigma_1-\hat{F}_1\sigma_S/\Gamma_S\right)^2/(\tau\sigma_0).\nonumber
\end{eqnarray}

The nonlinear $\mathbf{\Omega}_0$ eigenmode dispersion relation, can then be derived, by multiplying both sides of equation (\ref{eq:NL_test_DR_3}) with $\Phi^*_0$, noting  that $\mathscr{E}_S$ varies much slower than $|\Phi_0|^2$ and $|\Phi_1|^2$ in radial direction, and averaging over the radial length $1/(n_0q')\ll\delta\ll1/(n_Sq')$. One then has
\begin{eqnarray}
\left(\hat{\mathscr{E}}_0-\Delta_0 |A_1|^2-\chi_0\mathscr{E}_S |A_1|^2 \right)A_0 =-\frac{\hat{C}_0}{\mathscr{E}_S}|A_1|^2A_0,\label{eq:NL_test_DR_4}
\end{eqnarray}
in which, $\hat{\mathscr{E}}_0$ is the linear $\mathbf{\Omega}_0$  eigenmode dispersion relation, defined as $\hat{\mathscr{E}}_0=\int |\Phi_0|^2 \mathscr{E}_0 dr$. The coefficients, $\Delta_0$, $\chi_0$ and $\hat{C}_0$, corresponding respectively to nonlinear frequency shift, ion Compton scattering and shielded-ion scattering, are given as
\begin{eqnarray}
\Delta_0&=&\langle\langle \hat{\Lambda}^2(\hat{G}_1-\hat{H}_1)/b_0\rangle\rangle,\\
\chi_0&=& \langle\langle\hat{\Lambda}^2(\hat{G}_2-\hat{H}_2)/b_0\rangle\rangle,\\
\hat{C}_0&=&\langle\langle\hat{\Lambda}^2\hat{H}_3/b_0\rangle\rangle,
\end{eqnarray}
with
\begin{eqnarray}
\langle\langle\cdots\rangle\rangle\equiv\int (\cdots) |\Phi_0|^2|\Phi_1|^2dr\label{eq:averaging}
\end{eqnarray}
accounting for the contribution of TAE fine scale mode structures. Note that, because of this, equation (\ref{eq:averaging}) takes into account the selection rule on mode numbers that can be most effectively coupled via $\delta_1$; that is, the small normalized radial shift that accounts for the possible misalignment of TAM pump and decay modes.  The one-to-one correspondence to   $\Delta^{(2)}_{A_-}$, $\chi^{(2)}_{A_-}$ and $C_k$ of Ref. \citenum{LChenEPL2011} are straightforward.  $\chi_0$ can be further simplified, and yields
\begin{eqnarray}
\chi_0=\langle\langle \hat{\Lambda}^2\left(\hat{F}_2-\hat{F}^2_1/\Gamma_S\right)/(\tau b_0\sigma_0\Gamma_S)\rangle\rangle,
\end{eqnarray}
which is positive definite from Schwartz inequality \cite{LChenEPL2011}.

\subsection{Parametric decay instability}\label{sec:parametric_instability}

Equation (\ref{eq:NL_test_DR_4}) can be considered as the equation describing nonlinear parametric decay of a pump TAE ($\mathbf{\Omega}_1$) into  TAE/LKTAE  ($\mathbf{\Omega}_0$) and    ISW  ($\mathbf{\Omega}_S$) daughter waves  \cite{LChenEPL2011,ZQiuJPSCP2014}, and we immediately  obtain the nonlinear parametric dispersion relation
\begin{eqnarray}
\left(\hat{\mathscr{E}}_0-\Delta_0 |A_1|^2-\chi_0\mathscr{E}_S |A_1|^2 \right) =-\frac{\hat{C}_0}{\mathscr{E}_S}|A_1|^2,\label{eq:para_DR}
\end{eqnarray}
which can be solved for the condition of $\mathbf{\Omega}_1$ spontaneous decay.   The low- and high- $\beta$ regimes, will be discussed, respectively, in sections \ref{sec:low_beta} and \ref{sec:high_beta}.

\subsubsection{Low-$\beta$ limit: ion Compton scattering induced TAE cascading}
\label{sec:low_beta}

\begin{figure}
\includegraphics[width=3.0in]{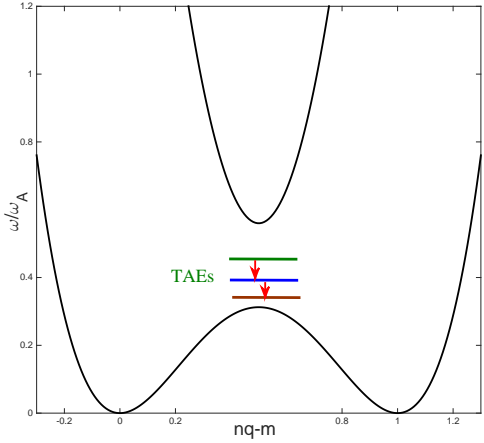}
\caption{TAE parametric decay in low-$\beta$ limit.} \label{Fig:TAE_ISW_TAE}
 \end{figure}

In the low-$\beta$ ($\beta\ll \epsilon^2$) limit, equation (\ref{eq:para_DR}) describes the pump TAE ($\mathbf{\Omega}_1$) decay into a sideband TAE ($\mathbf{\Omega}_0$) in the SAW continuum gap and an ISW ($\mathbf{\Omega}_s$), as shown in Fig. \ref{Fig:TAE_ISW_TAE}.  Since $\mathbf{\Omega}_S$ could be heavily ion Landau damped, depending on plasma parameters such as  $\tau$,    two parameter regimes with distinct decay mechanisms shall be discussed separately.

For typical tokamak parameters with $\tau\sim O(1)$,   $\mathbf{\Omega}_S$ is heavily ion Landau damped, and becomes a quasi-mode. One  obtains, from the imaginary part of equation (\ref{eq:para_DR}),
\begin{eqnarray}
\gamma+\gamma_0= \frac{|A_1|^2}{\partial_{\omega_0}\hat{\mathscr{E}}_{0,R}}\left(\frac{\hat{C}_0}{|\mathscr{E}_S|^2}+\chi_0 \right)\mathscr{E}_{S,I}.\label{eq:para_small_tao}
\end{eqnarray}
In deriving  equation (\ref{eq:para_small_tao}), $\hat{\mathscr{E}}_0\simeq i(\gamma+\gamma_0)\partial_{\omega_0}\hat{\mathscr{E}}_{0,R}$ expansion is taken, with  $\gamma_0\equiv -\hat{\mathscr{E}}_{0,I}/\partial_{\omega_0}\mathscr{E}_{0,R}$ being  the damping rates of $\mathbf{\Omega}_0$, and the subscript ``R" and ``I"  denoting real  and imaginary parts.
The two terms on the right hand side  of equation (\ref{eq:para_small_tao}) correspond to, respectively,  the shielded-ion and nonlinear ion Compton scatterings, and $\mathscr{E}_{S,I}$ is the imaginary part of $\mathbf{\Omega}_S$ dispersion function.  Noting that  $\hat{C}_0$ and $\chi_0$ are both positive definite, and that  $\mathscr{E}_{S,I}=\sqrt{\pi}\tau\Gamma_S\xi_S\exp(-\xi^2_S)$ with $\xi_S=(\omega_0-\omega_1)/(|k_{\parallel,S}v_{it}|)$, we then have, the parametric instability $\gamma>0$  requires $\omega_1>\omega_0$, i.e., the parametric decay can spontaneously  occur only when the pump TAE frequency is higher than that of the sideband. Thus,  the parametric decay process leads to, power transfer from higher to lower frequency part of the spectrum, i.e., downward spectrum cascading \cite{TSHahmPRL1995,LChenEPL2011}.   The threshold condition, is again, given by $\gamma=0$ for nonlinear drive via ion induced scattering to overcome  $\mathbf{\Omega}_0$ dissipation.

For $\tau\gg1$, on the other hand, $\mathbf{\Omega}_S$  is   weakly damped, and both $\mathbf{\Omega}_S$ and $\mathbf{\Omega}_0$ are normal modes of the system. Consequently the higher order terms $\Delta_0$ and $\chi_0$ can be neglected. The  resonant decay process can be analyzed following the  standard approach, and will be neglected here.

 \begin{figure}
\includegraphics[width=3.0in]{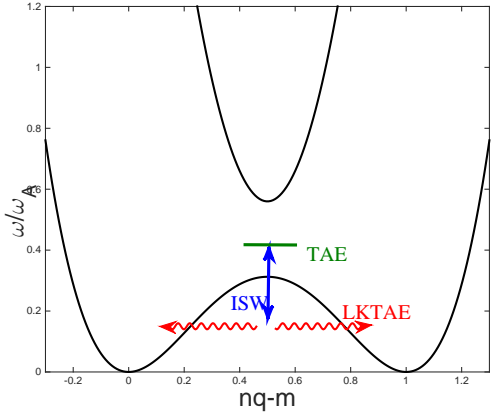}
\caption{TAE parametric decay in high-$\beta$ limit.} \label{Fig:TAE_ISW_LKTAE}
 \end{figure}

\subsubsection{High-$\beta$ limit: TAE decay into  LKTAE}\label{sec:high_beta}

In the high-$\beta$ ($\beta\geq\epsilon^2$) limit,  the sideband $\mathbf{\Omega}_0$ is a propagating LKTAE in the lower continuum, as shown in Fig. \ref{Fig:TAE_ISW_LKTAE}. Equations (\ref{eq:NL_quasimode}) and (\ref{eq:NL_test_DR_WKB_2}) can   be directly applied  here, while noting that  $\hat{\mathscr{E}}_0$ in equation (\ref{eq:NL_test_DR_WKB_2}) is now the LKTAE eigenmode dispersion relation, which    can be written as \cite{FZoncaPoP2014b,FZoncaPoP1996}
\begin{eqnarray}
\bar{\hat{\mathscr{E}}}_0=-\frac{\pi k^2_{\theta}\rho^2_i \omega^2_A}{2^{2\hat{\xi}+1} b_0\Gamma^2(\hat{\xi}+1/2)}\left[\frac{2\sqrt{2}\Gamma(\hat{\xi}+1/2)} {\hat{\alpha}\Gamma(\hat{\xi})} +\delta W_f\right],\nonumber
\end{eqnarray}
with the bar ($\bar{\cdots}$) denoting high-$\beta$ limit. Furthermore,  $\Gamma(\hat{\xi})$ and $\Gamma(\hat{\xi}+1/2)$  are Euler gamma functions,  $\hat{\xi}\equiv 1/4-\Gamma_+\Gamma_-/(4\sqrt{\Gamma_-\hat{s}^2\hat{\rho}^2_K})$, $\Gamma_{\pm}\equiv \omega^2/\omega^2_A(1\pm\epsilon_0)-1/4$, $\omega^2_A\equiv V^2_A/(q^2R^2_0)$, $\hat{\alpha}^2=1/(2\sqrt{\Gamma_-\hat{s}^2\hat{\rho}^2_K})$, $\hat{s}\equiv r\partial_r q/q$ is the magnetic shear,  $\delta W_f$ playing the role of a potential energy, and $\hat{\rho}^2_K\equiv (k^2_{\theta}\rho^2_i/2)\left[3/4+\tau (1-i\delta_e)\right]$ denotes  kinetic effects associated with finite ion Larmor radii and electron   Landau damping.

Note that, ISW frequency is higher for $T_e\gg T_i$ and in this ``high-$\beta$ limit",  resonant decay into  weakly ion Landau damped ISW is preferred. Neglecting higher order terms associated with $\mathbf{\Omega}_S$, equations (\ref{eq:NL_quasimode}) and (\ref{eq:NL_test_DR_WKB_2}) can be simplified as:
\begin{eqnarray}
\bar{\mathscr{E}}_S\delta\phi_S&=&i\frac{\hat{\Lambda}}{\omega_0}\sigma_0\sigma_1\delta\phi_0\delta\phi_{1^*},\label{eq:high_beta_ISW}\\
\bar{\hat{\mathscr{E}}}_0 \delta\phi_0&=& i\frac{\omega_0\hat{\Lambda}}{b_0}(\Gamma_S-\Gamma_1) \delta\phi_1\delta\phi_{S}.\label{eq:high_beta_LKTAE}
\end{eqnarray}
The parametric dispersion relation for a pump TAE ($\mathbf{\Omega}_1$) decaying into an ISW ($\mathbf{\Omega}_S$) and an LKTAE ($\mathbf{\Omega}_0$) is then
\begin{eqnarray}
\bar{\mathscr{E}}_S\bar{\hat{\mathscr{E}}}_0=-\langle\langle \sigma_0\sigma_1\hat{\Lambda}^2(\Gamma_S-\Gamma_1)/b_0\rangle\rangle |A_1|^2,
\end{eqnarray}
which can be solved following the standard procedure of resonant decay instabilities, and yields:
\begin{eqnarray}
(\gamma+\gamma_0)(\gamma+\gamma_S)=\frac{\langle\langle \sigma_0\sigma_1\hat{\Lambda}^2(\Gamma_S-\Gamma_1)/b_0\rangle\rangle|A_1|^2}{\partial_{\omega_0}\bar{\hat{\mathscr{E}}}_{0,R}\partial_{\omega_S}\bar{\mathscr{E}}_{S,R}}.\label{eq:para_high_beta}
\end{eqnarray}
Note that short radial scale averaging in equation (\ref{eq:para_high_beta}) introduces selection rules for the decay mode number, similar to the discussion following equation (\ref{eq:averaging}) above.

\section{TAE nonlinear saturation due to ion induced scattering}

As discussed in Sec. \ref{sec:parametric_instability}, spontaneous power transfer from $\mathbf{\Omega}_1$ to $\mathbf{\Omega}_0$   leads to the TAE scattering to the lower frequency fluctuation spectrum in the low-$\beta$ limit, and to LKTAE in the high-$\beta$ limit. In both cases, nonlinear saturation of the TAE fluctuation spectrum is eventually achieved. The TAE nonlinear saturation process and the resulting saturation level in low- and high-$\beta$ limit, are  analyzed in sections \ref{sec:spectrum_evolution} and \ref{sec:high_beta_saturation}.

\subsection{Low-$\beta$ limit: spectral transfer due to ion induced scattering} \label{sec:spectrum_evolution}

In a realistic burning plasma with typical  toroidal mode number $n\geq O(10)$ and finite $q$, many ($\sim O(n^2q)$) TAEs are excited by EPs with comparable linear growth rate \cite{LChenNF2007b}. Each TAE,  can thus, interact with the turbulence ``bath" of background TAEs, leading to nonlinear saturation and  spectrum transfer, as illustrated in Fig. \ref{Fig:cascading} and discussed in Ref. \citenum{TSHahmPRL1995}.  In the rest of section \ref{sec:spectrum_evolution}, we will investigate the TAE spectrum evolution, following closely the analysis of Ref. \citenum{TSHahmPRL1995}. The wave kinetic equation describing the spectrum evolution is derived in section \ref{sec:WKE}, which is then solved in section \ref{sec:WKE_solution} for the saturated spectrum and overall magnetic perturbation amplitude.

\subsubsection{Nonlinear wave-kinetic equation for  TAE spectrum evolution}\label{sec:WKE}

The above discussed nonlinear TAE spectral transfer, can be described by wave kinetic equation. Equation (\ref{eq:NL_test_DR_4}) describing the test TAE $\mathbf{\Omega}_0$ interacting with a background TAE $\mathbf{\Omega}_1$,  can be generalized  as
\begin{eqnarray}
 \hat{\mathscr{E}}_kA_k =\sum_{k_1}\left(\Delta_0 +\chi_0\mathscr{E}_S -\frac{\hat{C}_0}{\mathscr{E}_S}\right)|A_{k_1}|^2A_k,\label{eq:NL_test_DR_5}
\end{eqnarray}
with the subscript ``$k_1$" denoting background TAEs, and the summation over $k_1$ denoting   all the background TAEs within strong interaction region, i.e., counter-propagating and radially overlapping with $\mathbf{\Omega}_k$, and the frequency difference $|\omega_k-\omega_{k_1}|$ comparable with ion transit frequency ($|v_i/(qR_0)|$).  In equation (\ref{eq:NL_test_DR_5}), the $k$ subscript denotes a generic test TAE, and $\sum_{k_1}$ runs on all background modes, including $k$ itself. Furthermore, consistent with equation (\ref{eq:para_small_tao}), we have neglected the nonlinear frequency shift. Multiplying    equation (\ref{eq:NL_test_DR_5}) by $A_{k^*}$,  and taking the imaginary part, we then obtain  the wave kinetic equation describing  TAE   nonlinear evolution due to interaction with turbulence bath of  TAEs:
\begin{eqnarray}
&&\left(\partial_t-2\gamma_{L,k}\right)I_k\nonumber\\
&=&\frac{2}{\partial_{\omega_k}\hat{\mathscr{E}}_{k,R}}\sum_{k_1}\frac{1}{k^2_{\perp,1}}\left(\frac{\hat{C}}{|\mathscr{E}_S|^2}+\chi_0\right)\mathscr{E}_{S,i}I_{k_1}I_k,
\end{eqnarray}
with $I_k\equiv |\nabla_{\perp}A_k|^2$ and $\gamma_{L,k}\equiv -\hat{\mathscr{E}}_{k,I}/(\partial_{\omega_k}\hat{\mathscr{E}}_{k,R})$ being the linear growth/damping rate of $\mathbf{\Omega}_k$.

\begin{figure}
\includegraphics[width=3.5in]{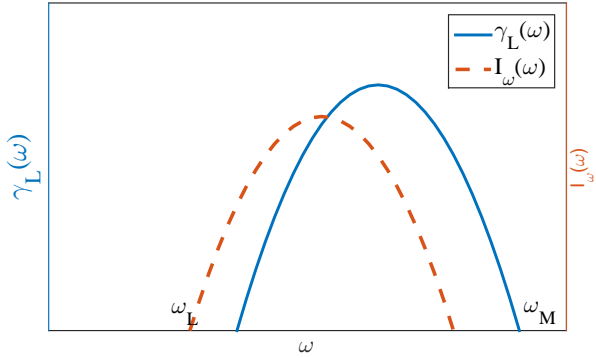}
\caption{TAE linear growth rate $\gamma_L$ and saturated spectrum $I_{\omega}$ dependence on $\omega$.} \label{Fig:cascading}
 \end{figure}

Denoting TAEs with their eigenfrequencies, i.e., $I_{k}\rightarrow I_{\omega}$,    the summation over ``$k_1$" can be replaced by integration over ``$\omega$", given many background TAEs within the strong interaction range with $\mathbf{\Omega}_{\omega}$ (continuum limit):
\begin{eqnarray}
\left(\partial_t-2\gamma_{L}(\omega)\right)I_{\omega}=\frac{2}{\partial_{\omega}\mathscr{E}_{\omega,R}}\int^{\omega_M}_{\omega_L} d\omega' V(\omega,\omega')I_{\omega'}I_{\omega}, \label{eq:spectrum_temporal_evolution}
\end{eqnarray}
with $I_{\omega}=\sum_k I_k\delta(\omega'-\omega_k)$ being the continuum version of $I_k$, $\omega_M$ being the highest frequency for TAE to be linearly unstable, $\omega_L$ being the lowest frequency for $I_{\omega_L}>0$ as shown in Fig. \ref{Fig:cascading},  and one has $\omega_M-\omega_L\simeq O(\epsilon)\omega_{T}$, comparable with the TAE gap width.  Furthermore $\omega_L$ is linearly stable, and is driven nonlinearly.   On the other hand,  the integration kernel $V(\omega,\omega')$ is defined as
\begin{eqnarray}
V(\omega,\omega')\equiv \frac{1}{k^2_{\perp,\omega'}}\left(\frac{\hat{C}}{|\mathscr{E}_S|^2}+\chi_0\right)\mathscr{E}_{S,i}.\nonumber
\end{eqnarray}

\subsubsection{Nonlinear saturation spectrum and magnetic fluctuation level}\label{sec:WKE_solution}

The nonlinear saturation condition can then be obtained from $\partial_tI_{\omega}=0$ as:
\begin{eqnarray}
\gamma_L(\omega)=-\frac{1}{\partial_{\omega}\mathscr{E}_{\omega,R}}\int^{\omega_M}_{\omega_L} d\omega' V(\omega,\omega')I_{\omega'}.
\end{eqnarray}

Noting that, for burning plasmas with most unstable TAEs characterized by toroidal mode number $n\gtrsim O(10)$, and that $|\omega_M-\omega_L|\gg|v_{it}/(qR_0)|$ for the process to be important,    $I_{\omega'}$ varies in $\omega'$ much slower than $V(\omega,\omega')$. Expanding $I_{\omega'}=I_{\omega}-\omega_S\partial_{\omega}I_{\omega}$,  the integral equation then becomes a differential equation, and we have
\begin{eqnarray}
\gamma_L(\omega)&=&-\frac{1}{\partial_{\omega}\mathscr{E}_{\omega,R}}\int^{\omega-\omega_L}_{\omega-\omega_M} d\omega_S V(\omega_S)\left(I_{\omega}-\omega_S\partial_{\omega}I_{\omega}\right)\nonumber\\
&=&-\frac{1}{\partial_{\omega}\mathscr{E}_{\omega,R}}\left[U_0I_{\omega}-U_1\partial_{\omega}I_{\omega}\right].\label{eq:spectrum_evolution}
\end{eqnarray}
Here, $U_0$ and $U_1$ are defined as, respectively,
\begin{eqnarray}
U_0&\equiv& \int^{\omega-\omega_L}_{\omega-\omega_M} d\omega_S V(\omega_S),\label{eq:u0}\\
U_1&\equiv& \int^{\omega-\omega_L}_{\omega-\omega_M} d\omega_S \omega_S V(\omega_S).\label{eq:u1}
\end{eqnarray}

For the ion Compton scattering process to be important, one requires $\omega_M-\omega_L\gg v_{it}/(qR_0)$, which corresponds to $\beta\ll\epsilon^2_0$. Noting that $V(\omega_S)\propto\mathscr{E}_{S,i}$ is an odd function of $\omega_S$  varying on the scale of $v_{it}/(qR_0)$,  $U_0$ becomes vanishingly small as $|\omega-\omega_L|, |\omega-\omega_M|\geq v_{it}/(qR_0)$. Equation (\ref{eq:spectrum_evolution}) can then be simplified, and yields
\begin{eqnarray}
I_{\omega}&\simeq&  \int^{\omega} d\omega \partial_{\omega}\mathscr{E}_{\omega,R}\frac{\gamma_L(\omega)}{U_1(\omega)}\nonumber\\
&\simeq& I_M(\omega_M)-\frac{1}{\overline{U_1}}\int^{\omega_M}_{\omega} \gamma_L(\omega)  \partial_{\omega}\mathscr{E}_{\omega,R} d\omega,\label{eq:u1_bar}
\end{eqnarray}
with $I_M(\omega_M)\equiv I_{\omega}(\omega=\omega_M)$. Note that, in equation (\ref{eq:u1_bar}), we have imposed boundary condition at $\omega=\omega_M$, and $\overline{U_1}$  is  moved out of the integration  due to the fact that $\mathscr{E}_{S,i}$ and, thus, $V(\omega_S)$ decays exponentially with $|\omega_S|$; thus, $U_1(\omega)$ is essentially constant under the integration sign. For    $|\omega-\omega_L|, |\omega-\omega_M|\gg v_{it}/(qR_0)$, the integral limits in equation (\ref{eq:u1}) can be replaced by $\pm\infty$, and we have
\begin{eqnarray}
\overline{U}_1&\simeq& \frac{1}{k^2_{\perp}}\left(\frac{\hat{C}}{|\mathscr{E}_S|^2}+\chi_0\right)\int^{\infty}_{-\infty}d\omega_S\omega_S\mathscr{E}_{S,i} \nonumber\\
&\simeq&\frac{\pi^{3/2}}{2k^2_{\perp}}\left(\frac{\hat{C}}{|\mathscr{E}_S|^2}+\chi_0\right)k^2_{\parallel,S}v^2_{it}.
\end{eqnarray}
The value of $I_{\omega}$ at $\omega_M$,  $I_M(\omega_M)$, on the other hand, can be determined  noting that for $|\omega-\omega_M|\ll |k_{\parallel,S}v_{it}|$, the lower and upper integral limits of equations (\ref{eq:u0}) and (\ref{eq:u1}), can be replaced by roughly, $0$ and $\infty$, and one has
\begin{eqnarray}
U_0(\omega_M)&\simeq&  \frac{1}{k^2_{\perp}}\left(\frac{\hat{C}}{|\mathscr{E}_S|^2}+\chi_0\right) \int^{\infty}_{0} d\omega_S\mathscr{E}_{S,i},\nonumber\\
&=&\overline{U_1}/(k_{\parallel,S}v_{it}),\nonumber\\
U_1(\omega_M)&\simeq&  \frac{1}{k^2_{\perp}}\left(\frac{\hat{C}}{|\mathscr{E}_S|^2}+\chi_0\right) \int^{\infty}_{0} d\omega_S\omega_S\mathscr{E}_{S,i}.\nonumber\\
&=&\overline{U_1}/2.\nonumber
\end{eqnarray}

$I_M(\omega_M)$ can then be derived from equation (\ref{eq:spectrum_evolution}), noting that $|U_0I_{\omega}/(U_1\partial_{\omega}I_{\omega})|\sim |(\omega_M-\omega_L)/(k_{\parallel,S}v_{it})|\gg1$, and one has
\begin{eqnarray}
I_{\omega}=\frac{2k_{\parallel,S}v_{it}\omega_M\gamma_L(\omega_M)}{\overline{U_1}}-\frac{1}{\overline{U_1}}\int^{\omega_M}_{\omega} \gamma_L(\omega) \partial_{\omega}\mathscr{E}_{\omega,R} d\omega.\nonumber
\end{eqnarray}

The overall TAE intensity at saturation, can be derived by integrating the intensity over the fluctuation population zone, and we have
\begin{eqnarray}
I_S&\equiv& \int^{\omega_M}_{\omega_L} I_{\omega} d\omega\simeq-\frac{\overline{\gamma_L}}{\overline{U_1}}\int^{\omega_M}_{\omega_L}(\omega-\omega_L)\partial_{\omega}\mathscr{E}_{\omega,R} d\omega\nonumber\\
&\simeq& \frac{\overline{\gamma_L}}{\overline{U_1}}\omega^3_T \epsilon^2_{eff},\label{eq:overall_intensity}
\end{eqnarray}
with $\epsilon_{eff}\equiv 1-\omega_M/\omega_L\sim O(\epsilon)$ following Ref. \citenum{TSHahmPRL1995}. In deriving equation (\ref{eq:overall_intensity}), we replaced the TAE linear growth rate $\gamma_L$ with its spectrum averaged value, $\gamma_L(\omega)\simeq\overline{\gamma_L}$, which is validated by the fact that, for burning plasma relevant parameter regimes,  a broad  TAE spectrum with comparable linear growth rate can be driven unstable \cite{LChenRMP2016,TWangPoP2018}.  In deriving the final expression of equation (\ref{eq:overall_intensity}), $\partial_{\omega}\mathscr{E}_{\omega,R}\sim -2\omega$ is used.
The contribution of $I_M$ is of order $\gamma_L(\omega_M)k_{\parallel,S}v_{it}/((\omega_M-\omega_L)\overline{\gamma_L})$ smaller than the other term, and is neglected.

The saturation level of the magnetic fluctuations, can then be obtained, noting $|\delta B_r|^2=|k_{\theta}\delta A_{\parallel}|^2=|ck_{\theta}k_{\parallel}/(\omega k_r)|^2 I_S$,
\begin{eqnarray}
|\delta B_r|^2\simeq \frac{c^2\epsilon^2\epsilon^2_{eff}}{2\pi^{3/2}}\frac{\omega_T\overline{\gamma}_Lk^2_r}{(\hat{C}/|\mathscr{E}_S|^2+\chi_0)\Omega^2_{ci}\rho^2_{it}} \label{eq:saturated_magnetic_perturbation}
\end{eqnarray}
with $|k_{\theta,T}/k_{r,T}|\simeq \epsilon$ for TAEs in the inertial layer assumed.  For $b\lesssim 1$, one has
\begin{eqnarray}
\hat{C}_0\sim \chi_0\sim \frac{c^2}{B^2_0}\tau  k^2_{r}k^2_{\theta} b\label{eq:Lambda}
\end{eqnarray}
and $|\mathscr{E}_S|\sim O(1)$, and thus,
\begin{eqnarray}
\left|\frac{\delta B_r}{B_0}\right|^2\sim \frac{\epsilon^4\epsilon^2_{eff}}{2\tau\pi^{3/2}}\frac{\overline{\gamma}_L}{\omega_T}  \frac{\omega^2_T/\Omega^2_{ci}}{k^4_{\theta}\rho^4_{it}}.
\label{eq:saturated_magnetic_perturbation_ratio}
\end{eqnarray}

The saturation level, $\delta B_r/B_0$,   is  smaller than the prediction of Ref. \citenum{TSHahmPRL1995} by  $\epsilon (\omega/\Omega_{ci})/( k^2_{\theta}\rho^2_{it})$ due to the enhanced coupling in the kinetic regime \cite{LChenEPL2011}. Noting that, for most unstable modes driven by EPs, one has typically, $k_{\theta}\rho_{d,E}\sim O(1)$ \footnote{Here, for simplicity of discussion while without loss of generality, well circulating EPs are assumed.}, which gives $k^2_{\theta}\rho^2_{it}\sim (T_i/T_E)/q^2$, with $\rho_{d,E}\sim(T_E/T_i)^{1/2} q \rho_{it}$ being the  EP magnetic drift orbit width and $T_E$ being the characteristic EP energy.  Thus, the saturation level predicted in the present work  applies as $T_i/(\epsilon T_Eq^2)\gg \omega/\Omega_{ci}$; while the expression given by Ref. \citenum{TSHahmPRL1995} can be used in the opposite limit. The saturation level given in equation (\ref{eq:saturated_magnetic_perturbation_ratio}), can then be simplified and yield the following scaling
\begin{eqnarray}
\left|\frac{\delta B_r}{B_0}\right|^2&\sim& \frac{m_i}{8\tau\pi^{3/2} e^2\mu_0}\frac{\overline{\gamma}_L}{\omega_T}\frac{T^2_E}{T^2_i}q^2N^{-1}_0\epsilon^6R^{-2}_0\nonumber\\
&\sim&1.2*10^{15}A_mq^2N^{-1}_0\epsilon^6R^{-2}_0\frac{T^2_E}{T^2_i}\frac{\overline{\gamma}_L}{\omega_T},\label{eq:low_beta_scaling}
\end{eqnarray}
with $A_m=m_i/m_p$ being the mass ratio of thermal ion to proton, and $N_0$ being the thermal plasma density.  For   typical burning plasmas parameters, the saturation level can be estimated as $|\delta B_r/B_0|\sim 10^{-4}-10^{-3}$. In obtaining the above saturation level, ITER like parameters are used, i.e., $B_0\sim 5 $ Tesla, $T_E\sim 3.5$ MeV, $T_i\sim 10$ KeV, $R_0\sim 6$ m, $N_0\sim 10^{20} m^{-3}$, $q\sim 3$, $\epsilon\sim 1/6-1/3$ and $\overline{\gamma}_L/\omega_T\sim 10^{-2}$.

\subsection{High-$\beta$ limit: TAE saturation via coupling to KTAE}\label{sec:high_beta_saturation}

In high-$\beta$ limit, the pump TAE decays into an ISW and a small scale LKTAE in the continuum. The resulting TAE saturation level, can be derived following the analysis of Ref. \citenum{ZQiuPRL2018}, where TAE decay into GAM and LKTAE is analyzed. For the simplicity of discussion, following the discussion of section \ref{sec:high_beta}, we assume $\mathbf{\Omega}_S$ is weakly ion Landau damped. The equation for the feedback of $\mathbf{\Omega}_0$ and $\mathbf{\Omega}_S$ to the unstable pump TAE $\mathbf{\Omega}_1$, is derived as
\begin{eqnarray}
\overline{\mathscr{E}}_1\delta\phi_1=i\frac{\omega_1}{b_1}\hat{\Lambda}(\Gamma_S-\Gamma_0)\delta\phi_0\delta\phi_{S^*}.\label{eq:high_beta_TAE}
\end{eqnarray}

The three-wave nonlinear dynamic equations can then be derived from equations (\ref{eq:high_beta_ISW}), (\ref{eq:high_beta_LKTAE}) and (\ref{eq:high_beta_TAE}) as
\begin{eqnarray}
(\partial_t+\gamma_S)A_S&=&\hat{\alpha}_SA_0A_{1^*},\\
(\partial_t+\gamma_0)A_0&=&\hat{\alpha}_0A_1A_S,\\
(\partial_t-\gamma_1)A_1&=&\hat{\alpha}_1A_0A_{S^*},
\end{eqnarray}
with $\gamma_1$ being the linear growth rate of the linearly unstable pump TAE due to, e.g.,  EP drive,  $\hat{\alpha}_S\equiv (\int   \Phi_0\Phi_{1^*}dr)^{-1}\int dr \Phi_0\Phi_{1^*}\hat{\Lambda} \sigma_0\sigma_1/(\omega_0\partial_{\omega_s}\bar{\mathscr{E}}_{S,R})$, $\hat{\alpha}_0\equiv  \omega_0\int dr |\Phi_0|^2|\Phi_1|^2 dr \hat{\Lambda}(\Gamma_S-\Gamma_1)/(b_0\partial_{\omega_0}\bar{\hat{\mathscr{E}}}_{0,R})$ and $\hat{\alpha}_1\equiv  \omega_1\int dr |\Phi_0|^2|\Phi_1|^2 dr \hat{\Lambda}(\Gamma_S-\Gamma_0)/(b_1\partial_{\omega_1}\bar{\mathscr{E}}_{1,R})$. The above coupled equations, describing the nonlinear evolution of the driven-dissipative system, may exhibit rich dynamics such as limit-cycle behaviors, period-doubling and route to chaos as possible indication of the existence of  attractors \cite{DRussellPRL1980}. In this work, focusing on  TAE nonlinear saturation and related transport, the TAE saturation level  can then be estimated from the fixed point solution, and one has
\begin{eqnarray}
|A_1|^2=\gamma_0\gamma_S/(\hat{\alpha}_S\hat{\alpha}_0).
\end{eqnarray}
Note that, the present analysis, assuming ISW being weakly ion Landau damped, can be readily generalized to ISW heavily ion Landau damped parameter regime,  by taking $\gamma_S\simeq v_{it}/(qR_0)$. The corresponding magnetic fluctuation amplitude, can then be estimated as
\begin{eqnarray}
\left|\frac{\delta B_r}{B_0}\right|^2&\simeq& \frac{c^2k^2_{\theta}k^2_{\parallel}}{\hat{\alpha}_S\hat{\alpha}_0B^2_0}\frac{\gamma_0\gamma_S}{\omega^2_1}.\label{eq:saturated_magnetic_perturbation_high_beta}
\end{eqnarray}
The magnitudes of $\hat{\alpha}_S$ and $\hat{\alpha}_0$, can be estimated in the $b\leq1$ limit, following the analysis of Ref. \citenum{ZQiuPRL2018}, and one obtains
\begin{eqnarray}
\left|\frac{\delta B_r}{B_0}\right|^2&\simeq&\frac{2\gamma_0\gamma_S}{\omega_0\omega_S}\frac{\epsilon^2k^2_{\parallel,0}}{k^2_{\theta,1}}\\
&\sim&6.5*10^{10}A_m\frac{\gamma_0\gamma_S}{\omega_0\omega_S}\epsilon^2R^{-2}_0B^{-2}_0T_E.\label{eq:scaling_high_beta}
\end{eqnarray}
The  TAE saturation level in the high-$\beta$ limit, can be estimated  as $|\delta B_r/B_0|\sim 10^{-4}$ for typical ITER-like parameters, assuming $\gamma_0/\omega_0\sim 10^{-2}$ and $\gamma_S/\omega_S\sim 1$.

The obtained TAE saturation level (and spectrum) given by equations (\ref{eq:saturated_magnetic_perturbation}) and (\ref{eq:saturated_magnetic_perturbation_high_beta}) in both low- and high-$\beta$ limit, can  be applied to derive the ion heating rate from ion Compton scattering rate (nonlinear Landau damping) \cite{TSHahmPST2015} and the EP transport coefficient \cite{LChenJGR1999} in the corresponding parameter regime. As an application,  presented in Sec. \ref{sec:transport} is  an estimation of EP transport coefficient   using quasilinear transport theory and  assuming well circulating EPs with relatively small magnetic drift orbit width. The ion heating rate is not derived here, while interested readers may readily derive it following the procedure of Ref. \citenum{TSHahmPST2015}, using the   saturated TAE amplitude given in equations (\ref{eq:saturated_magnetic_perturbation}) and (\ref{eq:saturated_magnetic_perturbation_high_beta}).

\section{Consequences on EP transport}
\label{sec:transport}

The EP transport coefficient, can be estimated from quasilinear transport theory \cite{LChenJGR1999}. For simplicity of discussion, well-circulating EPs with small drift orbit ($k_{\theta}\rho_{d,E}\lesssim1$) is assumed, while a more general approach is presented in Refs. \citenum{MFalessiPoP2018a,LChenSpringer2018,MFalessiPoP2018b}.  Considering transport time scale is much longer than the characteristic EP transit
time and spatial scale is much larger than resonant EP magnetic drift orbit width, the quasilinear equation for EP equilibrium distribution function evolution is \cite{LChenJGR1999}
\begin{eqnarray}
\partial_tF_{0,E}=-\frac{c}{B_0}\overline{\sum_{\mathbf{k}=\mathbf{k'}+\mathbf{k''}} \mathbf{\hat{b}}\cdot\mathbf{k''}\times\mathbf{k'}J_{k'}\delta L_{k'}\delta H_{k''}},\label{eq:quasi_linear_equation}
\end{eqnarray}
with $\overline{(\cdots)}$ denoting bounce averaging, $\mathbf{k}=\mathbf{k'}+\mathbf{k''}=k_Z\hat{\mathbf{r}}$ selecting phase space zonal structure \cite{FZoncaNJP2015} modulations in the radial direction, and $\delta H$ being the linear EP response to $\mathbf{k''}$, consistent with the quasilinear ordering.  For well circulating EPs,  $\delta H_{k}$ can be derived by transforming into drift orbit center coordinates, and one obtains \footnote{For the derivation of equation (\ref{eq:linear_EP_response}), interested readers may refer to Ref. \citenum{ZQiuPoP2016} and references therein.}
\begin{eqnarray}
\delta H_{k}&=&-\frac{e}{m}Q_kF_0 J_k\delta L_k\sum_{l,p}\frac{J_l(\hat{\lambda}_k)J_p(\hat{\lambda}_k) e^{-i(l-p)(\theta-\theta_{0r})}}{\omega_k-k_{\parallel}v_{\parallel}+l\omega_{tr}},\nonumber\\
&&\label{eq:linear_EP_response}
\end{eqnarray}
with $\hat{\lambda}_k=k_{\perp}\hat{v}_d/\omega_{tr}$ denoting finite drift orbit width effects, and $\theta_{0r}\equiv \tan^{-1}(k_r/k_{\theta})$. Substituting equation (\ref{eq:linear_EP_response}) into equation (\ref{eq:quasi_linear_equation}),
one then has,
\begin{eqnarray}
&&\partial_tF_{0,E} = i\frac{c}{B_0}\frac{e}{m}k_{\theta}\frac{\partial}{\partial r}\left[J^2_k|\delta L_k|^2\sum_l J^2_l(\hat{\lambda}_k)\right.\nonumber\\
&&\left.\times\left(\frac{1}{\omega_k-k_{\parallel}v_{\parallel}+l\omega_{tr}}-\frac{1}{\omega^*_k-k_{\parallel}v_{\parallel}+l\omega_{tr}}\right)Q_kF_{0}\right].\nonumber
\end{eqnarray}
Noting that
\begin{eqnarray}
&&\left(\frac{1}{\omega_k-k_{\parallel,k}v_{\parallel}+l\omega_{tr}}-\frac{1}{\omega^*_k-k_{\parallel,k}v_{\parallel}+l\omega_{tr}}\right)\nonumber\\
&=&-2i\pi\delta (\omega_k-k_{\parallel}v_{\parallel}+l\omega_{tr})\nonumber
\end{eqnarray}
and that $\delta  L_k\simeq (1-k_{\parallel}v_{\parallel}/\omega_k)\delta\phi_k$, one then has
\begin{eqnarray}
\partial_t N_{0,E}\simeq - \partial_rD_{Res}\partial_r N_{0,E},
\end{eqnarray}
with $N_{0,E}$ being the equilibrium EP density, and the resonant EP radial diffusion rate given as
\begin{eqnarray}
D_{Res}&\equiv&\left\langle2\pi\sum_l |\delta V_{Er,l}|^2J^2_l(\hat{\lambda}_k)\delta (\omega-k_{\parallel}v_{\parallel}+l\omega_{tr}) \frac{F_0}{N_{0,E}}\right\rangle,\nonumber\\
&&\label{eq:EP_ql_coefficient}
\end{eqnarray}
and $|\delta V_{Er,l}|^2\equiv c^2k^2_{\theta}J^2_k|\delta\phi_k|^2  l^2\omega^2_{tr}/(B^2_0\omega^2_k)$ being the resonant EP radial electric-field drift velocity. For EPs with small magnetic drift orbits, $k_{\theta}\rho_{d,E}\lesssim1$, $l=\pm1$ transit harmonic resonances dominate, and the EP radial transport  coefficient is very similar to that describing zero frequency zonal flow generation by EP driven TAEs \cite{ZQiuPoP2016} (equation (10) therein), with the underlying mechanism that zonal structures   are linearly un-damped  strucures related to nonlinear equilibria. It is also straightforward  to find out that, $J^2_l(\hat{\lambda}_k)\delta (\omega-k_{\parallel}v_{\parallel}+l\omega_{tr})$ is proportional to the linear growth rate of  TAE, and this reflects the fundamental property that resonant particle transport and wave-particle power transfer (resonant excitation) are intimately related \cite{LChenRMP2016}.

The  circulating EP transport coefficient induced by the saturated TAE spectrum in the short wavelength $k^2_{\theta}\rho^2_i/\epsilon\gg\omega/\Omega_{ci}$ limit, can be derived by substituting equations (\ref{eq:saturated_magnetic_perturbation})  or (\ref{eq:saturated_magnetic_perturbation_high_beta}) into equation (\ref{eq:EP_ql_coefficient}). Noting $|\delta\phi|^2=\omega^2\delta B^2_r/(c^2k^2_{\theta}k^2_{\parallel})$, one then obtains
\begin{eqnarray}
D_{Res}&\simeq&\frac{1}{4}\frac{V_A}{k_{\parallel,0}}\left|\frac{\delta B_r}{B_0}\right|^2.\label{eq:EP_diffusion_rate}
\end{eqnarray}
In deriving the EP diffusion rate, equation (\ref{eq:EP_diffusion_rate}),   resonant EP transit time, $\omega_{tr,Res}^{-1}$, is taken as the de-correlation time. Substituting equation (\ref{eq:low_beta_scaling}) into (\ref{eq:EP_diffusion_rate}),  one then has, the scaling law for TAE induced EP diffusion rate in the low-$\beta$ limit
\begin{eqnarray}
D_{Res}\sim 1.3*10^{31}A^{1/2}_m\epsilon^6q^3N^{-3/2}_0R^{-1}_0\frac{T^2_E}{T^2_i}\frac{\overline{\gamma}_L}{\omega_T}.\label{eq:EP_diffusion_scaling}
\end{eqnarray}
For ITER-like parameters, the circulating EP diffusion rate can be estimated  as $D_{Res}\sim 1 - 10^2  m^2/s$, for $\epsilon\sim 1/6 - 1/3$.
Note that, this coefficient is valid for circulating particles in the lower-$\beta$ limit, as   ion induced  scattering  is the dominant mechanism for  TAE nonlinear saturation. The corresponding result for the higher-$\beta$ limit, meanwhile, can be obtained similarly from equation  (\ref{eq:EP_diffusion_rate}) with $\delta B_r$ given by equation (\ref{eq:saturated_magnetic_perturbation_high_beta}), and one has $D_{Res}\sim 1 m^2/s$.
For potential predictive applications, calibration using results from large scale  simulations  \cite{WZhangPRL2008} and/or test particle simulations \cite{ZFengPoP2013} is required, and this will be carried out in a future publication.

\section{Conclusions and Discussions}\label{sec:summary}

In conclusion, the  TAE saturation  in the burning plasma related short wavelength limit ($k^2_{\theta}\rho^2_i/\epsilon\gg \omega/\Omega_{ci}$) is analyzed, using nonlinear gyrokinetic equation.  In the low-$\beta$  limit, with $\beta\ll\epsilon^2$, a TAE may decay into another TAE with lower frequency due to ion induced scattering.  The nonlinear equation  describing a test TAE nonlinear evolution due to  interacting with a background TAE is derived, which is then generalized to including all the background TAEs that are strongly interacting with the test TAE  for burning plasmas with most unstable TAEs characterized by typically $n\gtrsim 10$ \cite{LChenRMP2016,IPBNF1999,AFasoliNF2007,TWangPoP2018}. It is shown that  the damping of the generated ISW due to ion induced scattering plays a key role  in the nonlinear decay process, and the spontaneous decay requires that the secondary generated TAEs have a lower frequency, leading to the downward  TAE spectral transfer and finally  saturation due to enhanced coupling to SAW continuum. The wave kinetic equation describing TAE spectral transfer is derived from the imaginary part of the nonlinear TAE equation,  which is then solved for the nonlinear saturated spectrum.  In the high-$\beta$ limit, with $\beta\gg\epsilon^2$, the TAE may directely decay   into a lower KTAE in the continuum, and the corresponding parametric dispersion relation as well as result TAE saturation level is also derived.  The related EP transport coefficient  is derived using quasilinear transport theory \cite{LChenJGR1999}, assuming, as illustration, well circulating EPs with small drift orbits,   that the transport time scale is slower than particle transit time, and that the corresponding spatial scale is longer than EP magnetic drift orbit.

For the processes discussed in this paper to occur and dominate over other mechanisms,  several constraint on plasma parameters are required. First, $k^2_{\theta}\rho^2_{it}/\epsilon >\omega/\Omega_{ci}$  for the nonlinear coupling in the kinetic regime to dominate over that due to parallel ponderomotive force. Second, for the process discussed in Sec. \ref{sec:low_beta} to occur,  $\beta\ll \epsilon^2$  is required for the high frequency secondary SAW mode due to this nonlinear ion induced scattering process to be a  gap TAE. This $\beta\ll \epsilon^2$  regime is also assumed for solving the wave-kinetic equation for the saturated TAE spectrum. Meanwhile, for the process discussed in Sec. \ref{sec:high_beta} to be dominant,  $\beta\gg \epsilon^2$ is required for   the   high frequency sideband to be a LKTAE  with the frequency lower than the lower accumulational point frequency of toroidicity induced gap.

As a final remark, the nonlinear ion induced scattering discussed here, has a cross-section  comparable to other  processes in the short wavelength $k^2_{\theta}\rho^2_{it}/\epsilon >\omega/\Omega_{ci}$ limit,  e.g., ZFZS generation \cite{LChenPRL2012} and/or  decaying into a GAM and a kinetic TAE (KTAE) \cite{ZQiuPRL2018}. Thus, the TAE saturation  can be quite sensitive to the threshold condition of different channels; i.e., it depends on the considered plasma parameter regime and, typically, multiple nonlinear physics processes are responsible for it.

\section*{Acknowledgements}

This work is supported by   the National Key R\&D Program of China  under Grant No.  2017YFE0301900,
the National Science Foundation of China under grant Nos.  11575157 and 11875233,      EUROfusion Consortium
under grant agreement No. 633053 and US DoE GRANTs.

\section*{References}

\providecommand{\newblock}{}


\begin{thebibliography}{10}
\expandafter\ifx\csname url\endcsname\relax
  \def\url#1{{\tt #1}}\fi
\expandafter\ifx\csname urlprefix\endcsname\relax\def\urlprefix{URL }\fi
\providecommand{\eprint}[2][]{\url{#2}}

\bibitem{KTomabechiNF1991}
Tomabechi K, Gilleland J, Sokolov Y, Toschi R and Team I 1991 {\em Nuclear
  Fusion\/} {\bf 31} 1135

\bibitem{YKolesnichenkoVAE1967}
Kolesnichenko Y~I 1967 {\em At. Energ\/} {\bf 23} 289

\bibitem{AMikhailovskiiSPJ1975}
Mikhailovskii A 1975 {\em Zh. Eksp. Teor. Fiz\/} {\bf 68} 25

\bibitem{MRosenbluthPRL1975}
Rosenbluth M and Rutherford P 1975 {\em Phys. Rev. Lett.\/} {\bf 34} 1428

\bibitem{LChenPoP1994}
Chen L 1994 {\em Physics of Plasmas\/} {\bf 1} 1519--1522

\bibitem{LChenRMP2016}
Chen L and Zonca F 2016 {\em Review of Modern Physics\/} {\bf 88} 015008

\bibitem{IPBNF1999}
ITER Physics Expert Group~on Energetic~Particles Heating and Current Drive  and ITER Physics Basis Editors 
  1999 {\em Nuclear Fusion\/} {\bf 39} 2471

\bibitem{AFasoliNF2007}
Fasoli A, Gormenzano C, Berk H, Breizman B, Briguglio S, Darrow D, Gorelenkov
  N, Heidbrink W, Jaun A, Konovalov S, Nazikian R, Noterdaeme J~M, Sharapov S,
  Shinohara K, Testa D, Tobita K, Todo Y, Vlad G and Zonca F 2007 {\em Nuclear
  Fusion\/} {\bf 47} S264

\bibitem{RDingNF2015}
Ding R, Pitts R, Borodin D, Carpentier S, Ding F, Gong X, Guo H, Kirschner A,
  Kocan M, Li J, Luo G~N, Mao H, Qian J, Stangeby P, Wampler W, Wang H and Wang
  W 2015 {\em Nuclear Fusion\/} {\bf 55} 023013

\bibitem{CZChengAP1985}
Cheng C, Chen L and Chance M 1985 {\em Ann. Phys.\/} {\bf 161} 21

\bibitem{LChenJGR1999}
Chen L 1999 {\em Journal of Geophysical Research: Space Physics\/} {\bf 104}
  2421--2427 ISSN 2156-2202

\bibitem{YTodoPoP1995}
Todo Y, Sato T, Watanabe K, Watanabe T and Horiuchi R 1995 {\em Physics of
  Plasmas\/} {\bf 2} 2711--2716

\bibitem{JLangPoP2010}
Lang J, Fu G~Y and Chen Y 2010 {\em Physics of Plasmas\/} {\bf 17} 042309

\bibitem{JZhuPoP2013}
Zhu J, Fu G and Ma Z 2013 {\em Physics of Plasmas\/} {\bf 20} 072508

\bibitem{SBriguglioPoP2014}
Briguglio S, Wang X, Zonca F, Vlad G, Fogaccia G, Di~Troia C and Fusco V 2014
  {\em Physics of Plasmas\/} {\bf 21} 112301

\bibitem{JZhuNF2014}
Zhu J, Ma Z and Fu G 2014 {\em Nuclear Fusion\/} {\bf 54} 123020

\bibitem{DSpongPoP1994}
Spong D, Carreras B and Hedrick C 1994 {\em Physics of plasmas\/} {\bf 1}
  1503--1510

\bibitem{TSHahmPRL1995}
Hahm T~S and Chen L 1995 {\em Phys. Rev. Lett.\/} {\bf 74}(2) 266--269

\bibitem{FZoncaPRL1995}
Zonca F, Romanelli F, Vlad G and Kar C 1995 {\em Phys. Rev. Lett.\/} {\bf 74}
  698

\bibitem{LChenPPCF1998}
Chen L, Zonca F, Santoro R and Hu G 1998 {\em Plasma physics and controlled
  fusion\/} {\bf 40} 1823

\bibitem{YTodoNF2010}
Todo Y, Berk H and Breizman B 2010 {\em Nuclear Fusion\/} {\bf 50} 084016

\bibitem{LChenPRL2012}
Chen L and Zonca F 2012 {\em Phys. Rev. Lett.\/} {\bf 109}(14) 145002

\bibitem{ZQiuEPL2013}
Qiu Z, Chen L and Zonca F 2013 {\em Europhysics Letters\/} {\bf 101} 35001

\bibitem{ZQiuPoP2016}
Qiu Z, Chen L and Zonca F 2016 {\em Physics of Plasmas (1994-present)\/} {\bf
  23} 090702

\bibitem{ZQiuNF2017}
Qiu Z, Chen L and Zonca F 2017 {\em Nuclear Fusion\/} {\bf 57} 056017

\bibitem{TSHahmPST2015}
Hahm T~S 2015 {\em Plasma Science and Technology\/} {\bf 17} 534

\bibitem{ZQiuPRL2018}
Qiu Z, Chen L, Zonca F and Chen W 2018 {\em Phys. Rev. Lett.\/} {\bf 120}
  135001

\bibitem{NFischPRL1992}
Fisch N~J and Rax J~M 1992 {\em Phys. Rev. Lett.\/} {\bf 69}(4) 612--615

\bibitem{NFischNF1994}
Fisch N and Herrmann M 1994 {\em Nuclear Fusion\/} {\bf 34} 1541

\bibitem{LChenEPL2011}
Chen L and Zonca F 2011 {\em Europhysics Letters\/} {\bf 96} 35001

\bibitem{EFriemanPoF1982}
Frieman E~A and Chen L 1982 {\em Physics of Fluids\/} {\bf 25} 502--508

\bibitem{FZoncaPoP1996}
Zonca F and Chen L 1996 {\em Physics of Plasmas\/} {\bf 3} 323--343

\bibitem{LChenPS1995}
Chen L and Zonca F 1995 {\em Physica Scripta\/} {\bf 1995} 81

\bibitem{JConnorPRL1978}
Connor J, Hastie R and Taylor J 1978 {\em Phys. Rev. Lett.\/} {\bf 40} 396

\bibitem{ZQiuNF2016}
Qiu Z, Chen L and Zonca F 2016 {\em Nuclear Fusion\/} {\bf 56} 106013

\bibitem{GFuPoFB1989}
Fu G~Y and Van~Dam J~W 1989 {\em Physics of Fluids B\/} {\bf 1} 1949--1952

\bibitem{FZoncaPoFB1993}
Zonca F and Chen L 1993 {\em Physics of Fluids B: Plasma Physics\/} {\bf 5}
  3668--3690

\bibitem{ZQiuJPSCP2014}
{Qiu} Z, {Chen} L and {Zonca} F 2014 {\em JPS Conference Proceedings\/} {\bf 1}
  015007

\bibitem{FZoncaPoP2014b}
Zonca F and Chen L 2014 {\em Physics of Plasmas\/} {\bf 21} 072121

\bibitem{LChenNF2007b}
Chen L and Zonca F 2007 {\em Nuclear Fusion\/} {\bf 47} S727

\bibitem{TWangPoP2018}
Wang T, Qiu Z, Zonca F, Briguglio S, Fogaccia G, Vlad G and Wang X 2018 {\em
  Physics of Plasmas\/} {\bf 25} 062509

\bibitem{DRussellPRL1980}
Russell D~A, Hanson J~D and Ott E 1980 {\em Phys. Rev. Lett.\/} {\bf 45}(14)
  1175--1178

\bibitem{MFalessiPoP2018a}
Falessi M~V and Zonca F 2018 {\em Physics of Plasmas\/} {\bf 25} 032306

\bibitem{LChenSpringer2018}
Chen L and Zonca F 2018 ``Physics of alfv\'en waves and energetic particles", ``Springer Monograph" in preparation

\bibitem{MFalessiPoP2018b}
Falessi M and Zonca F 2018 ``Transport theory of phase space zonal structures", submitted to {\em Physics of Plasmas}

\bibitem{FZoncaNJP2015}
Zonca F, Chen L, Briguglio S, Fogaccia G, Vlad G and Wang X 2015 {\em New
  Journal of Physics\/} {\bf 17} 013052

\bibitem{WZhangPRL2008}
Zhang W, Lin Z and Chen L 2008 {\em Phys. Rev. Lett.\/} {\bf 101} 095001

\bibitem{ZFengPoP2013}
Feng Z, Qiu Z and Sheng Z 2013 {\em Physics of Plasmas (1994-present)\/} {\bf
  20} 122309

\end{thebibliography}

\end{document}